\documentstyle[12pt,aaspp4]{article}
\lefthead{KALOGERA AND BAYM}
\righthead{MAXIMUM NEUTRON STAR MASS}
\begin{document}
\begin{center}
To appear in {\em The Astrophysical Journal Letters}
\end{center}
\vspace{1.cm}
\title{The Maximum Mass of a Neutron Star}
\author{Vassiliki Kalogera}
\affil{Astronomy Department,
       University of Illinois at Urbana-Champaign,\\
	1002 W. Green St., Urbana, IL 61801,\\
	vicky@astro.uiuc.edu}
\begin{center}
and
\end{center}
\author{Gordon Baym}
\affil{Physics Department,
	University of Illinois at Urbana-Champaign,\\
	1110 W. Green St., Urbana, IL 61801,\\
      gbaym@uiuc.edu}
\newcommand{\m}{maximum neutron-star mass }
\begin{abstract}

    Observational identification of black holes as members of binary systems
requires the knowledge of the upper limit on the gravitational mass of a
neutron star.  We use modern equations of state for neutron star matter,
fitted to experimental nucleon-nucleon scattering data and the properties of
light nuclei, to calculate, within the framework of Rhoades \& Ruffini (1974),
the minimum upper limit on a neutron star mass.  Regarding the equation of
state as valid up to twice nuclear matter saturation density, $\rho_{nm}$, we
obtain a secure upper bound on the neutron star mass equal to
$2.9$\,M$_\odot$.  We also find that in order to reach the lowest possible
upper bound of 2.2\,M$_\odot$, we need understand the physical properties of
neutron matter up to a density of $\sim 4\rho_{nm}$.

\end{abstract}

\section{INTRODUCTION}

    The identification of a compact object as a black hole requires not only
an accurate observational estimate of its mass but also knowledge of the
maximum gravitational mass of a neutron star for stability against collapse
into a black hole.  A growing number of measured mass functions for X-ray
binaries containing compact objects accreting mass from their companions has
appeared in the literature in the last years.  Knowledge of characteristics of
the companion stars of these objects allows one to impose lower limits on
their masses and to begin to identify black holes observationally (for recent
reviews see Cowley\markcite{C94} 1994 and Tanaka \& Lewin\markcite{T95} 1995).
Determination of neutron star masses and the range they cover also provides
significant information related to the mechanism responsible for neutron star
formation and possible effects of the evolutionary history of their
progenitors on their masses.

    The calculation of an accurate maximum neutron-star mass strongly depends
on our knowledge of the equation of state of neutron matter up to very high
densities, $\sim 5\times 10^{15}$\,g\,cm$^{-3}$.  Modern equations of state
are derived from accurate numerical solutions of the quantum-mechanical
nuclear many-body problem using two and three-body potentials fitted to
experimental nucleon-nucleon (NN) scattering data as well the properties of
light nuclei, which give information in the region of nuclear matter
saturation density, $\rho_{nm} \sim 2.7\times 10^{14}$\,g\,cm$^{-3}$.  See
Wiringa, Fiks, \& Fabrocini\markcite{W88} (1988) (hereafter WFF88) for a
description of such derivations and further references.  The equations of
state can be safely regarded as accurate up to $\sim 2\rho_{nm}$, and possibly
extended up to $\sim 4\rho_{nm}$, while the study of neutron star structure
demands an extrapolation towards much higher densities, where the properties
of matter remain rather uncertain.  In particular, less-well known intrinsic
three-nucleon forces become more important with increasing density, as do
further hadronic and eventually quark degrees of freedom (reviewed in
Baym\markcite{B95} (1995)).

    A powerful approach to the problem is to derive least upper bounds to the
maximum allowed gravitational mass of a neutron star by using the properties
of neutron matter at density ranges where they can be accurately predicted,
and imposing a minimum number of constraints at densities exceeding a higher
fiducial density, $\rho_0$, e.g., subluminal sound velocity and thermodynamic
stability.  In this way, uncertainties related to the properties of
high-density neutron matter can be circumvented in obtaining a secure upper
limit to the maximum allowed mass of a neutron star.  A number of authors have
approached the problem at different levels of completeness and for different
sets of fundamental assumptions.  Hartle\markcite{H78} (1978) offers a
thorough and detailed review of past work.  Oppenheimer \&
Volkoff\markcite{O39} (1939) were the first to suggest that the mass of a
stable neutron star becomes maximum for the stiffest possible equation of
state that is consistent with fundamental physical constraints.  Rhoades \&
Ruffini\markcite{R74} (1974), following this approach, were the first to
prove this suggestion rigorously.  Using a variational method in which they
constrained the equation of state to have subluminal sound velocity and be
stable against microscopic collapse, they proved that, in the regions where it
is uncertain, the equation of state that produces the maximum neutron star
mass is the one for which the sound speed is equal to the speed of light.  As
a result they found a maximum neutron-star mass $\simeq 3.2$\,M$_\odot$
assuming uncertainty in the equation of state above a fiducial density
$\rho_0=4.6\times 10^{14}$ \,g\,cm$^{-3}$.

    In this paper we revisit the issue of the maximum neutron-star mass,
employing recent equations of state for neutron matter of WFF88 in the
framework of the variational analysis of Rhoades \& Ruffini (1974).  In the
next section we describe our calculation, the resulting neutron star models,
and the maximum neutron-star mass as a function of the fiducial density,
$\rho_0$, up to which one believes the modern neutron star matter equations of
state to be reliable.  A principal result of this study is to show the
dependence of the maximum neutron mass on current uncertainties in the
equation of state.  In \S\,3, we compare our results to those of earlier
studies and discuss their implications regarding the observational
identification of black holes in binary systems.

\section{MAXIMUM NEUTRON STAR MASS}

    To calculate a least upper bound on the maximum mass of a neutron star
stable against gravitational collapse we follow the initial argument of
Rhoades \& Ruffini (1974) and consider a neutron star to be divided in two
parts:  (i) an outer envelope of which the mass is calculated based on a
specific neutron matter equation of state, and (ii) an inner core whose mass
we extremize.  The interface between these two regions is at a specified
fiducial density, $\rho_0$.  The set of fundamental constraints, independent
of the detailed physical properties of neutron matter, imposed on the equation
of state of the inner core are the following:  (i) the mass density, $\rho$, is
non-negative, i.e., gravity is attractive; (ii) the pressure, $P$, at zero
temperature is a function of $\rho$ only, i.e., neutron matter is a fluid;
(iii) $dP/d\rho \geq 0$, so that the zero-frequency sound speed of neutron
matter $(dP/d\rho)^{1/2}$ is real and matter is stable against microscopic
collapse; (iv) the sound speed does not exceed the speed of light, i.e.,
$dP/d\rho\leq c^2$, hence signals cannot be superluminal and causality is
satisfied.\footnote{The connection between the zero frequency sound velocity
being greater than the speed of light and violation of causality, while
physically plausible, is a tricky question, due to the presence of the
frequency dependence of the sound velocity and sound wave damping.  We are not
aware of a general proof yet that the ground state of matter must obey
$dP/d\rho\leq c^2$.  See Hartle (1978) for further discussion of this issue.}
As Ruffini \& Rhoades show, under these conditions the mass of the neutron
star becomes maximum for the stiffest possible equation of state, one for
which the sound speed, $c_s$, is equal to the speed of light:
\begin{equation}
c_s^2 = \frac{dP}{d\rho} = c^2.
\end{equation}

    To evaluate the maximum mass of a neutron star, we adopt, at densities
below $\rho_0$, the recent equations of state presented by WFF88, which
represent the most complete microscopic calculations to now that use available
low-energy nuclear data.  WFF88 derive the equation of state for two different
Hamiltonians, using the Argonne v$_{14}$ (AV14) and the Urbana v$_{14}$ (UV14)
two-nucleon potentials. In contrast to mean field calculations fit to saturation
properties, the two potentials are directly fitted with high accuracy to 
elastic nucleon-nucleon scattering
data and the properties of the deuteron. The two-nucleon potentials are 
supplemented with the Urbana VII
(UVII) three-nucleon potential constructed to provide the saturation of bulk
nuclear matter at 16 MeV binding energy at $\rho_{nm}$, and to fit the
properties of light nuclei. The results obtained by WFF88 represent the best 
microscopic
equation of state for dense matter constrained by nucleon-nucleon scattering
data. The two variations of the equation of state are
tabulated in Table V of WFF88.  For densities lower than $2.5\times
10^{14}$\,g\,cm$^{-3}$, we use the equation of state given by Baym, Pethick,
\& Sutherland\markcite{B71} (1971) for convenience, since the details of the
equation of state at these low densities do not affect our results.  Above the
fiducial density, $\rho_0$, we continue the WFF88 equations of state from
their values at $\rho_0$ assuming $c_s = c$.

    We note that for the two variations of the equation of state used above
$\rho = 2.5\times 10^{14}$\,g\,cm$^{-3}$, the sound speed exceeds the speed of
light at very high densities, $\rho \gtrsim 1.6\times 10^{15}$\,g\,cm$^{-3}$.
This violation of causality arises from the form of the three-body forces used
(WFF88).  Essentially, a three-body force that produces an energy density
$E_3\sim n^3$, as in the simply UVII interaction, where $n$ is the baryon
density, leads to a contribution to the pressure $P_3 = 2E_3$, which tends
increases $c_s$ above $c$ at very high density.  A more careful treatment of
the momentum dependence of the three-body force is required to avoid this
problem.  Since subluminal sound velocity is one of the constraints imposed on
the equation of state of the inner core, the results for fiducial densities
exceeding $1.6\times 10^{15}$\,g\,cm$^{-3}$ become suspect; however, as we see
below, the maximum mass as a function of $\rho_0$ reaches a plateau before
this point.

    Having specified the equation of state and the fiducial density, $\rho_0$,
we then calculate a series of neutron star models for a range of central
densities, $\rho_c$, by directly integrating the general relativistic equation
of hydrostatic equilibrium, the Tolman-Oppenheimer-Volkoff (TOV) equation:
\begin{equation}
\frac{dP}{dr}=-\frac{G}{r^2}\left[\rho(r)+\frac{P(r)}{c^2}\right]~
\left[m(r)+4\pi r^3\frac{P(r)}{c^2}\right]~
\left[1-2G\frac{m(r)}{rc^2}\right]^{-1},
\end{equation}
out to the radius $R$ of the neutron star, where $P(R)$ falls to zero; in
equation (2)
\begin{equation}
m(r)=\int_0^r 4\pi r^{'2} \rho(r^{'} )dr^{'}.
\end{equation}

    The resulting neutron star masses as a function of central density,
$\rho_c$, for different fiducial densities, $\rho_0$, are shown in Figure 1
for the AV14 plus UVII potential.  As one would expect, as $\rho_0$ decreases
and the stiffest part of the equation of state becomes more dominant, the
neutron star mass at specified central density increases.  Using these
``maximal" models we calculate the maximum neutron-star mass as a function of
the fiducial density, $\rho_0$, shown in Figure 2 for the two variations of
the equation of state used below $\rho_0$.  At low densities the results are
well approximated by:
\begin{equation}
M_{max}=6.7\,M_\odot~\left(\frac{\rho_0}{10^{14}\mbox{g\,cm}^3}\right)^{-1/2}.
\end{equation}
At densities higher than $\simeq 10^{15}$\,g\,cm$^{-3}$, $M_{max}$
approaches a limiting value $\simeq 2.2$\,M$_\odot$ independent of $\rho_0$.
The reason is that at high fiducial densities the mass of the inner core
becomes negligible compared to that of the outer envelope and consequently,
the upper bound is governed by the maximum mass of the envelope, which is
independent of $\rho_0$.  Note that this limiting value is reached before the
WFF88 equations of state become superluminal.

    In Figure 3 we show the radii of the maximal neutron star models of mass
1.4M$_\odot$ and 1.8 M$_\odot$, as a function of the fiducial density for the
two equations of state of WFF88.  Here the UV14 plus UVII equation of state
(dotted lines) produces slightly larger stars at higher $\rho_0$ than the AV14
plus UVII equation of state (solid lines).  The dependence of the neutron star
radius on mass is weak for $\rho_0 \gtrsim 10^{15}$\,g\,cm$^{-3}$.
At low densities the radius of a 1.4M$_\odot$ neutron star is well approximated
by:
\begin{equation}
R_{1.4}=21.2\,{\rm km}~
\left(\frac{\rho_0}{10^{14}\mbox{g\,cm}^3}\right)^{-0.35}.
\end{equation}

\section{DISCUSSION}

    We have taken into account recent neutron star matter equations of state
to determine a new  upper bound on the mass of a neutron star stable
against gravitational collapse.  Our results agree with the qualitative
behavior of $M_{max}$ as a function of the fiducial density, $\rho_0$, found
in earlier studies (see Hartle 1978).  The normalization of our empirical
relation at low densities is also consistent with earlier results.  At higher
densities $M_{max}$ becomes independent of $\rho_0$, and approaches a plateau
value $\simeq 2.2$\,M$_\odot$, higher than previously found.  Effects of
rotation become important only if the neutron stars rotate close to break-up.
Friedman \& Ipser\markcite{F87} (1987) have studied the effect of
uniform rotation (differential rotation is efficiently damped;
Hegyi\markcite{H77} 1977), and found that maximum rotation can cause an
increase of $\sim 25\%$ on the maximum neutron-star mass .

    The lowest value of $M_{max}$ is determined by the maximum value of
$\rho_0$ up to which one can be confident of the WFF88 equation of state.
Their results can be regarded as valid up to $\rho_0=2\rho_{nm}$, at which
density $M_{max}=2.9$\,M$_\odot$.  We consider this value to be our safest
estimate.  If we take their results as valid up to $\rho_0 \sim 4\rho_{nm}$,
$M_{max}$ is further reduced to $\simeq 2.2$\,M$_\odot$.  Further increase of
$\rho_0$ does not affect the value of $M_{max}$, which remains constant at
$\simeq 2.2$\,M$_\odot$.  The appearance of this plateau is significant
because it indicates that in order to have an accurate estimate of the lowest
possible upper bound on the neutron star mass, we need only understand the
detailed properties of neutron matter up to a density $\sim 4\rho_{nm}$, which
is a feasible extrapolation from low energy nuclear data.

    The updated upper bound of 2.9\,M$_\odot$ on the neutron star mass can be
used to discriminate between neutron stars and black holes based on
measurements of the mass function of the companion to the compact object.  The
mass function depends on the inclination, $i$, i.e., the angle between the
line of sight and the normal to the orbital plane, and the mass of the
companion.  Although the inclination of the plane of the orbit is generally
not known, its maximum value ($i=90^\circ$) can be used along with a lower
limit to the mass of the companion to set a lower limit on the mass of the
compact object.  If this limit exceeds the value of $M_{max}$, one has strong
evidence that the compact object is a black hole.  In this way, nine black
holes candidates have been identified so far.  The lower limits of their
masses lie in the range from 3.1\,M$_\odot$ to 6\,M$_\odot$.  Since the value
of $M_{max}=2.9$\,M$_\odot$ we report here is substantially lower than the one
used to date (3.2\,M$_\odot$), it is quite probable that more compact objects
can be identified as black holes.

\acknowledgements

    It is a pleasure to thank Arya Akmal for useful discussions and Dimitrios
Psaltis for helpful comments and for a critical reading of the manuscript.
This work was supported in part by NSF Grants AST92-18074 and PHY94-21309.

\newpage

\begin{center}
\large
{\bf Figure Captions}
\normalsize
\end{center}

    \figcaption{Neutron star mass, $M$, as a function of central density,
$\rho_c$, for three different values of the fiducial density, $\rho_0$.  At
densities below $\rho_0$ the equation of state based on the AV14 plus UVII
potential is used.  Vertical lines lie at constant $\rho_0 =
\rho_{nm}$, $2\rho_{nm}$, and $4\rho_{nm}$, where nuclear matter density
$\rho_{nm}=2.7\times 10^{14}$\,g\,cm$^{-3}$.}

    \figcaption{Maximum neutron star mass, $M_{max}$, as a function of the
fiducial density, $\rho_0$, for the two variations of the WFF88 equation of
state:  AV14 plus UVII potential (solid line) and UV14 plus UVII potential
(dotted line).  Vertical dashed lines lie at constant $\rho_0 = \rho_{nm}$,
$2\rho_{nm}$, and $4\rho_{nm}$.}

    \figcaption{Neutron star radius, $R$, as a function of the fiducial
density, $\rho_o$.  Curves are plotted for two different neutron star masses,
$M_{NS}=1.4$\,M$_\odot$ (thick lines) and $M_{NS}=1.8$\,M$_\odot$ (thin
lines), and for the two variations of the WFF88 equation of state:  AV14 plus
UVII potential (solid lines) and UV14 plus UVII potential (dotted lines).
Vertical dashed lines lie at constant $\rho_0 = \rho_{nm}$, $2\rho_{nm}$, and
$4\rho_{nm}$.}


\begin{references}

\reference{B95} Baym, G. 1995, Nucl. Phys. A, 590, 233c

\reference{B71} Baym, G., Pethick, C., \& Sutherland, P. 1971, \apj , 170, 299

\reference{C94} Cowley, A. P. 1994, in The Evolution of X-Ray Binaries, ed.
S. S. Holt \& C. S. Day (New York: AIP), 45

\reference{F87} Friedman, J. L., \& Ipser, J. R. 1987, \apj , 314, 594

\reference{H78} Hartle, J. B. 1978, Phys. Rep., 46, 201

\reference{H77} Hegyi, D. J. 1977, \apj , 217, 244

\reference{O39} Oppenheimer, J. R., \& Volkoff, G. M. 1939, Phys. Rev., 55, 374

\reference{R74} Rhoades, C. E. Jr., \& Ruffini, R. 1974, \prl , 32, 324

\reference{T95} Tanaka, Y., \& Lewin, W. H. G. 1995, in X-Ray Binaries, ed.
W. H. G. Lewin, J. van Paradijs, \& E. P. J. van den Heuvel (Cambridge: 
Cambridge Univ. Press), 126

\reference{W88}	Wiringa, R. B., Fiks, V., Fabrocini, A. 1988, \prc , 38, 1010

\end{references}
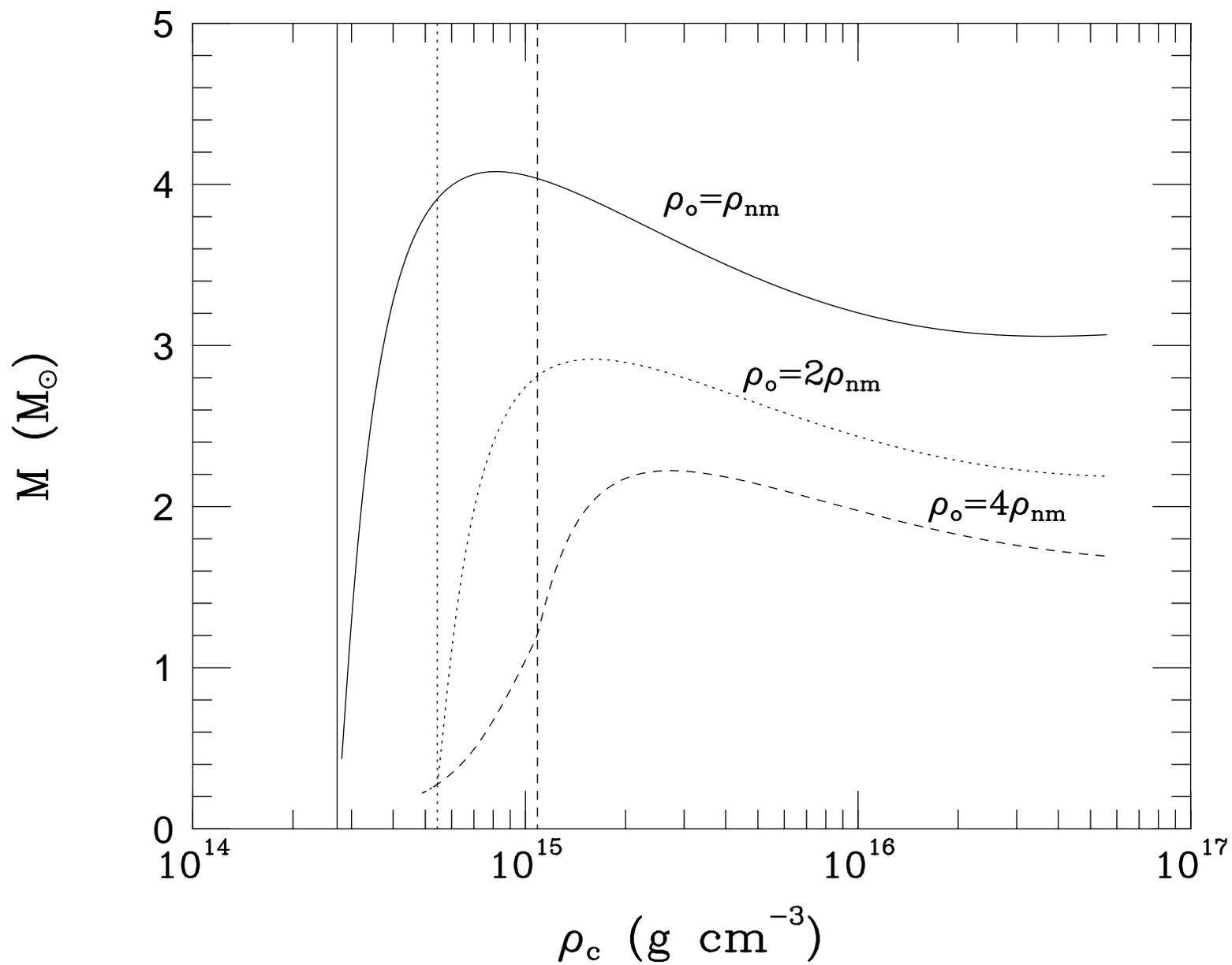
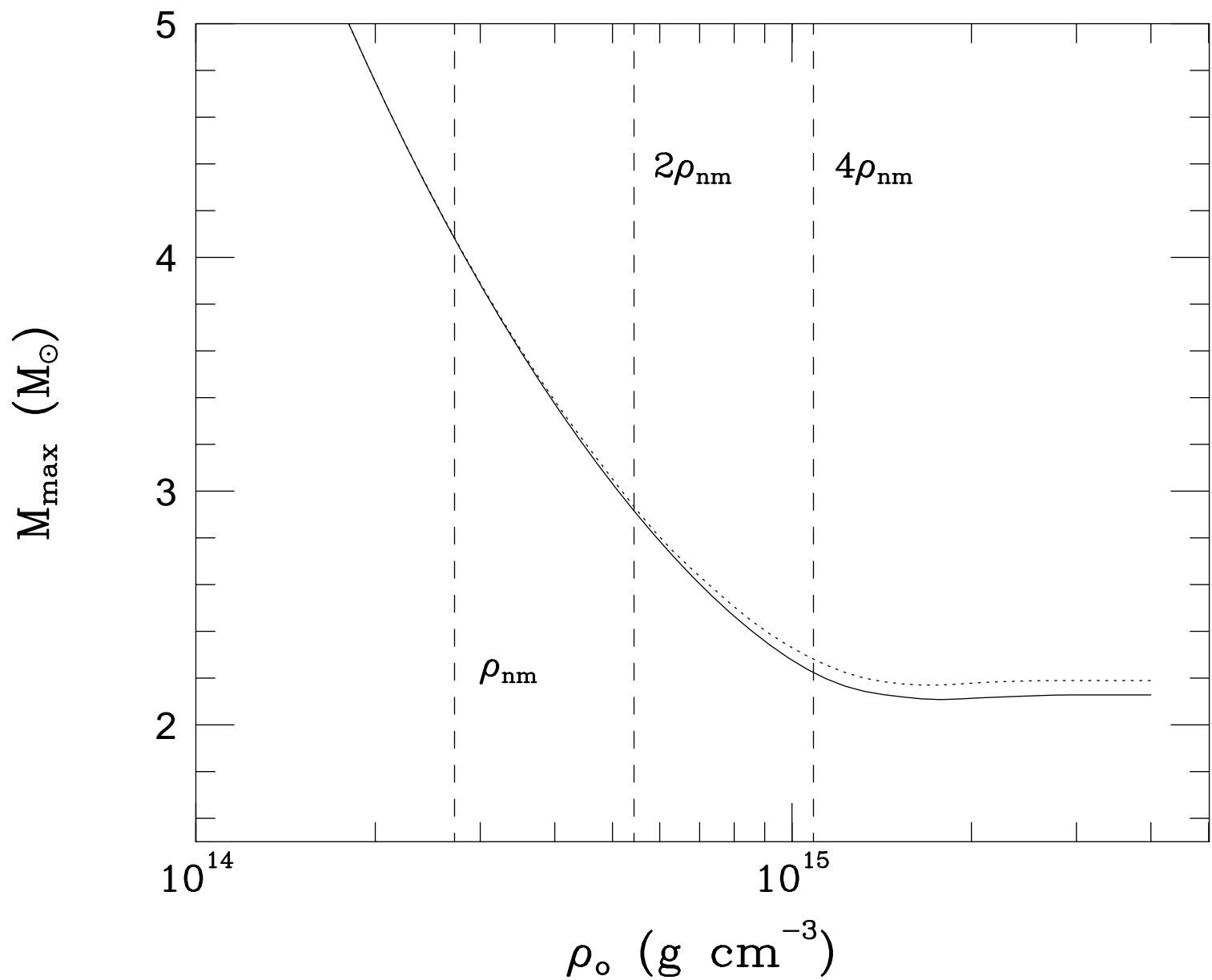
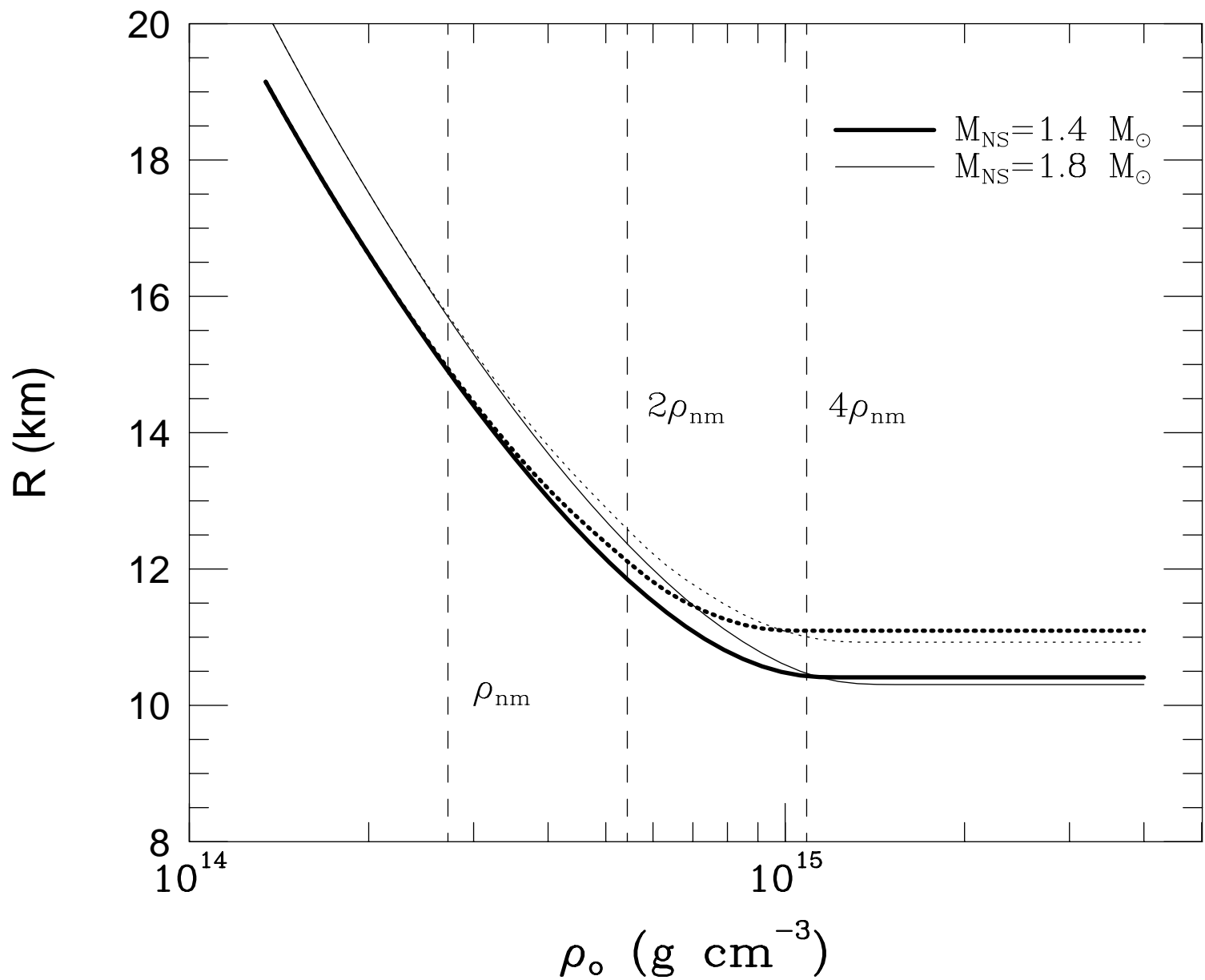
\end{document}